\documentclass{article}
\usepackage{latexsym}
\usepackage{graphicx}
  
\begin{document}
\Large
\noindent 
\textbf{A new solution of the Schr\"odinger equation}

\large
\begin{center}
\renewcommand{\thefootnote}{\fnsymbol{footnote}}
Yoshio Kishi\footnote[1]{Yoshio Kishi is the person to correspond with. 3-25-19 Chidori Ota-ku, Tokyo 146-0083, Japan. E-mail : yoshio.kishi@dream.com} and SeiIchiro Umehara

\end{center}

\normalsize
\begin{center}
\textbf{Abstract}
\end{center}

We obtained a new solution of Schr\"odinger equation by the method of Euclidean approach (Wick rotation). This is a wave motion which is fluctuating.

\begin{center}
\textbf{1. Introduction}
\end{center}

In the Feynman's path integral, the solution of the Schr\"odinger equation is expressed as\\
$\displaystyle \psi\left(x,t\right)=\int Dx\exp \left[{i\over\hbar }\int_{t_{0}}^{t}{1\over 2}mv^2 dt'\right]\left[\exp\left(-{i\over \hbar}\int_{t_{0}}^{t}V\left(x,t'\right)dt'\right)\psi\left(x_{0},t_{0}\right)\right]$.\\
And\\
$\displaystyle \int Dx\exp \left[{i\over\hbar }\int{1\over 2}mv^2 dt\right]=\int Dx\exp \left[-{1\over\hbar }\int{1\over 2}m{\left(dx\right)^2\over d\left(it\right)}\right]$\\
corresponds to the normal distribution after replacing $it\rightarrow t$ , so wave motion function can be written as\\
$\displaystyle \psi\left(x,t\right)=E\left[\exp\left(-{i\over \hbar}\int_{t_{0}}^{t}V\left(x,t'\right)dt'\right)\psi\left(x_{0},t_{0}\right)\right]$.\\
$E\left[\cdots\right]$ means the expected value is taken. This is an expression of so-called Feynman-Kac formula.
The purpose of our study is to discover the correct wave motion function that had been concealed by the operation of taking the expected value.

\begin{center}
\textbf{2. Discussion}
\end{center}

We try to obtain a new solution of Schr\"odinger equation\\
$\displaystyle i\hbar{\partial \psi\left(x,t\right) \over \partial t}=-{\hbar^2 \over 2m}{\partial^2 \psi\left(x,t\right) \over \partial x^2}+V\left(x,t\right)\psi\left(x,t\right)$\hfill{}(1)\\
by the technique of the Euclidean approach (Wick rotation). The Euclidean approach is one of the techniques of the quantum electrodynamics to come and go in the quantum mechanics and the statistical mechanics by doing a $it\rightarrow t$ replacement.\\
(1) becomes\\
$\displaystyle -{\partial \psi\left(x,t\right) \over \partial t}+{\hbar \over 2m}{\partial^2 \psi\left(x,t\right) \over \partial x^2}={V\left(x,t\right)\over\hbar}\psi\left(x,t\right)$\hfill{}(2)\\
after using the Euclidean approach ($it\rightarrow t$) and dividing both sides by $\hbar$.

The normal distribution\\
$\displaystyle \int Dx\exp \left[-{1\over\hbar }\int{1\over 2}m{\left(dx\right)^2\over dt}\right]$\\
means that $x$ follows the stochastic process of\\
$\displaystyle dx=\sqrt{\hbar\over m}dW^\ast\left(t\right)$ (Refer to Appendix A)\hfill{}(3).\\
Here,\\
$dW^\ast\left(t\right)=\sqrt{-dt}\xi$ $\left(dt<0\right)\cdots$Standard Brownian motion (conjugate Wiener process)\\
$\xi\cdots$Standard regular random variable.\\
(By the replacement of $it\rightarrow \tau$, path of integration $t$ is converted into imaginary number time like $\left[t_{1},t_{2}\right]\rightarrow \left[\tau_{1}/i,\tau_{2}/i\right]\rightarrow \left[-i\tau_{1},-i\tau_{2}\right]$. On the other hand, because real time $\tau$ is $\tau_{1}-\tau_{2}>0$, the direction of time increasing becomes contrary to the direction of integration ($d\tau=\tau_{2}-\tau_{1}<0$). In a word, it becomes $dt\rightarrow -id\tau$ for $it \rightarrow \tau$.)\\
If Ito's lemma (Refer to Appendix B) is used, a function $\psi\left(x,t\right)$ that has variable $x$ and $t$  follows the stochastic process of\\
$\displaystyle d\psi\left(x,t\right)=\left({\partial \psi\left(x,t\right) \over \partial t}-{1 \over 2}{\partial^2 \psi\left(x,t\right) \over \partial x^2}\left\{\sqrt {\hbar\over m}  \right\}^2 \right)dt+{\partial \psi\left(x,t\right) \over \partial x}\sqrt {\hbar\over m}dW^\ast\left(t\right)$\hfill{}(4).\\
If (2) is substituted for (4), it becomes\\
$\displaystyle d\psi\left(x,t\right)=-{V\left(x,t\right)\over\hbar}\psi\left(x,t\right)dt+{\partial \psi\left(x,t\right) \over \partial x}\sqrt {\hbar\over m}dW^\ast\left(t\right)$\hfill{}(5).\\
This is Ornstein-Uhlenbeck process. So the solution is obtained as follows$^{1}$.\\
$\displaystyle \psi\left(x,t\right)=\exp\left(-{1\over \hbar}\int_{t_{0}}^{t}V\left(x,t'\right)dt'\right)\left[\right.\psi\left(x_{0},t_{0}\right)$\\
$\displaystyle +\int_{t_{0}}^{t}\exp\left({1\over \hbar}\int_{t_{0}}^{u}V\left(x,t'\right)dt'\right){\partial \psi\left(x,u\right) \over \partial x}\sqrt {\hbar\over m}dW^\ast\left(u\right)\left.\right]$\\
or\\
$\displaystyle \psi\left(x,t\right)=\exp\left(-{1\over \hbar}\int_{t_{0}}^{t}V\left(x,t'\right)dt'\right)\psi\left(x_{0},t_{0}\right)$\\
$\displaystyle +\int_{t_{0}}^{t}\exp\left(-{1\over \hbar}\int_{u}^{t}V\left(x,t'\right)dt'\right){\partial \psi\left(x,u\right) \over \partial x}\sqrt {\hbar\over m}dW^\ast\left(u\right)$\hfill{}(6).\\
Clause 2 of the right side shows fluctuation. Then we can obtain a new solution of Schr\"odinger equation (fluctuating wave function) by reverse Euclidean approach ($t\rightarrow it$).\\
$\displaystyle \psi\left(x,t\right)=\exp\left(-{i\over \hbar}\int_{t_{0}}^{t}V\left(x,t'\right)dt'\right)\left[\right.\psi\left(x_{0},t_{0}\right)$\\
$\displaystyle +\int_{t_{0}}^{t}\exp\left({i\over \hbar}\int_{t_{0}}^{u}V\left(x,t'\right)dt'\right){\partial \psi\left(x,u\right) \over \partial x}\sqrt {\hbar\over m}\sqrt{-idu}\xi\left.\right]$\\
$\displaystyle =\exp\left(-{i\over \hbar}\int_{t_{0}}^{t}V\left(x,t'\right)dt'\right)\left[\right.\psi\left(x_{0},t_{0}\right)$\\
$\displaystyle +\exp\left(-{\pi\over 4}i\right)\int_{t_{0}}^{t}\exp\left({i\over \hbar}\int_{t_{0}}^{u}V\left(x,t'\right)dt'\right){\partial \psi\left(x,u\right) \over \partial x}\sqrt {\hbar\over m}\sqrt{du}\xi\left.\right]$\\
or\\
$\displaystyle \psi\left(x,t\right)=\exp\left(-{i\over \hbar}\int_{t_{0}}^{t}V\left(x,t'\right)dt'\right)\psi\left(x_{0},t_{0}\right)$\\
$\displaystyle +\exp\left(-{\pi\over 4}i\right)\int_{t_{0}}^{t}\exp\left(-{i\over \hbar}\int_{u}^{t}V\left(x,t'\right)dt'\right){\partial \psi\left(x,u\right) \over \partial x}\sqrt {\hbar\over m}dW\left(u\right)$\hfill{}(7).\\
This is the correct expression of a fluctuating wave motion. Clause 1 shows the appearance that the principal ingredient of the wave develops at time by exponential with potential in the shoulder. Only this clause 1 is considered in a present quantum theory. The effect of fluctuation is added by clause 2.\\
\hfill{}\\
We will confirm the fluctuation disappears from the expression when the expected value of this new solution (fluctuating solution) of the Schr\"odinger equation is taken.\\
After taking expected value of the right side of (6) by\\
$\displaystyle \int Dx\exp \left[-{1\over\hbar }\int{1\over 2}m{\left(dx\right)^2\over dt}\right]$,\\
(6) becomes\\
$\displaystyle \psi\left(x,t\right)=\int Dx\exp \left[-{1\over\hbar }\int{1\over 2}mv^2 dt\right]\left[\exp\left(-{1\over \hbar}\int_{t_{0}}^{t}V\left(x,t'\right)dt'\right)\psi\left(x_{0},t_{0}\right)\right]$\hfill{}(8)\\
because clause 2 of the right side of (6) is $0$ by Ito's integral.\\
It is understood that clause 2 of (6) disappears and the fluctuation has disappeared on 
the expression. And replace $t\rightarrow it$ in (8), then (8) becomes\\
$\displaystyle \psi\left(x,t\right)=\int Dx\exp \left[{i\over\hbar }\int{1\over 2}mv^2 dt\right]\left[\exp\left(-{i\over \hbar}\int_{t_{0}}^{t}V\left(x,t'\right)dt'\right)\psi\left(x_{0},t_{0}\right)\right]$\hfill{}(9).\\
It is corresponding to Feynman's path integral formula (Lagrangian Path Integrals).\\
In Hamiltonian Path Integrals\\
$\displaystyle \psi\left(x,t\right)=\int DxDp_{x}\exp\left({i\over\hbar}\int \left[p_{x}\dot x-{p_{x}^{2}\over 2m}-V\left(x,t\right)\right]dt\right)\psi\left(x_{0},t_{0}\right)$\\
$\displaystyle =\int DxDp_{x}\exp\left({i\over\hbar}\int \left[-{1 \over 2m}\left(p_{x}-m{dx\over dt}\right)^2+{1\over 2}m\left({dx \over dt}\right)^2-V\left(x,t\right)\right]dt\right)\psi\left(x_{0},t_{0}\right)$\\
, if the Gauss integration concerning the momentum is executed, it becomes Lagrangian Path Integrals.
When a kinetic energy paragraph of Hamiltonian Path Integrals\\
$\displaystyle \int Dp_{x}\exp\left({i\over\hbar}\int \left[-{1 \over 2m}\left(p_{x}-m{dx\over dt}\right)^2\right]dt\right)$\\ 
is compared with the normal distribution function, it is meant that the momentum fluctuates like\\
$\displaystyle dp_{x}\cong \sqrt{\hbar m \over dt}$.\\
(3) means coordinates fluctuates like\\
$\displaystyle dx\cong \sqrt{{\hbar \over m} dt}$.\\
So, It becomes\\
$\displaystyle \Delta x \Delta p_{x} \cong \hbar$\\
from this two. It is a so-called uncertainty principle.

\begin{center}
\textbf{3. The image of fluctuating wave function}
\end{center}

If (8) is differentiated by $x$ , it becomes\\
$\displaystyle {\partial \psi\left(x,t\right) \over \partial x}=\left\{-{1\over\hbar}{V\left(x,t\right)dt\over dx}\right\}\psi\left(x,t\right)$\hfill{}(10).\\
So after (10) is substituted for (5), it becomes\\
$\displaystyle d\psi\left(x,t\right)=-{V\left(x,t\right)\over\hbar}\psi\left(x,t\right)dt+\left\{-{1\over\hbar}{V\left(x,t\right)dt\over dx} \right\}\psi\left(x,t\right)\sqrt {\hbar\over m}dW^\ast\left(t\right)$\hfill{}(11).\\
If both sides is divided by $\psi\left(x,t\right)$ , it becomes\\
$\displaystyle {d\psi\left(x,t\right)\over\psi\left(x,t\right)}=-{V\left(x,t\right)\over\hbar}dt+\left\{-{1\over\hbar}{V\left(x,t\right)dt\over dx}\right\}\sqrt {\hbar\over m}dW^\ast\left(t\right)$\hfill{}(12).\\
$\displaystyle {d\psi\left(x,t\right)\over\psi\left(x,t\right)}=d\log\left[\psi\left(x,t\right)\right]=\log\left[\psi\left(x,t\right)\right]-\log\left[\psi\left(x\left(t-dt\right),t-dt\right)\right]=\log{\psi\left(x,t\right)\over \psi\left(x\left(t-dt\right),t-dt\right)}$\\
, So (12) becomes\\
$\displaystyle \log{\psi\left(x,t\right)\over \psi\left(x\left(t-dt\right),t-dt\right)}=-{V\left(x,t\right)\over\hbar}dt+\left\{-{1\over\hbar}{V\left(x,t\right)dt\over dx}\right\}\sqrt {\hbar\over m}dW^\ast\left(t\right)$\hfill{}(13).\\
Therefore, $\psi\left(x,t\right)$ becomes\\
$\displaystyle \psi\left(x,t\right)=\exp\left(-{1\over\hbar}\left[V\left(x,t\right)dt+\left\{{V\left(x,t\right)dt\over dx}\right\}\sqrt {\hbar\over m}dW^\ast\left(t\right)\right]\right)\psi\left(x\left(t-dt\right),t-dt\right)$\\
$\displaystyle =\exp\left(-{1\over\hbar}V\left(x,t\right)\left[dt+{dt\over dx}\sqrt {\hbar\over m}dW^\ast\left(t\right)\right]\right)\psi\left(x\left(t-dt\right),t-dt\right)$\\
$\displaystyle =\exp\left(-{1\over\hbar}V\left(x,t\right)\left[dt+{1\over v_{x}}\sqrt {\hbar\over m}dW^\ast\left(t\right)\right]\right)\psi\left(x\left(t-dt\right),t-dt\right)$\hfill{}(14).\\
A Brownian motion paragraph appears to the shoulder of exponential, and, as a 
result, the wave function fluctuates.

Considering that the potential is a function of also $y$ like $V\left(x\right)=V\left(x,y\right)$ , the discussion when the electron is turning to the vertical direction ($y$) against the incidence direction ($x$) receiving power is as follows.\\
$\displaystyle \psi\left(x,t\right)=\exp\left({1\over\hbar}p_{y}\cdot dy \right)\psi\left(x\left(t-dt\right),t-dt\right)$\hfill{}(15).\\
$\displaystyle p_{y}=-{\partial V\left(x,y\right)\over\partial y}\left[dt+{1\over v_{x}}\sqrt {\hbar\over m}dW^\ast\left(t\right)\right]$\hfill{}(16).\\
It is understood that the momentum in the direction of $y$ fluctuates by the Brownian 
motion.

When we replace $t$ with $it$ in (14), it becomes\\
$\displaystyle \psi\left(x,t\right)=\exp\left(-{i\over\hbar}V\left(x,t\right)\left[dt+{1\over v_{x}}\sqrt {\hbar\over m}\sqrt{-idt}\xi\right]\right)\psi\left(x\left(t-dt\right),t-dt\right)$\\
$\displaystyle =\exp\left(-{i\over\hbar}V\left(x,t\right)\left[dt+\left(1-i\right){1\over v_{x}}\sqrt {\hbar\over 2m}\sqrt{dt}\xi\right]\right)\psi\left(x\left(t-dt\right),t-dt\right)$\\
$\displaystyle =\exp\left(-{1\over\hbar}V\left(x,t\right)\left[{1\over v_{x}}\sqrt {\hbar\over 2m}dW\left(t\right)\right]\right)     \exp\left(-{i\over\hbar}V\left(x,t\right)\left[dt+{1\over v_{x}}\sqrt {\hbar\over 2m}dW\left(t\right)\right]\right)\psi\left(x\left(t-dt\right),t-dt\right)$\hfill{}(17).\\

When we replace $t$ with $it$ in (15) and (16), it becomes\\
$\displaystyle \psi\left(x,t\right)=\exp\left(-{1\over\hbar}{\partial V\left(x,t\right) \over \partial y}\left[{1\over v_{x}}\sqrt {\hbar\over 2m}dW\left(t\right)\right]\cdot dy \right)    \exp\left({i\over\hbar}p_{y}\cdot dy \right)\psi\left(x\left(t-dt\right),t-dt\right)$\hfill{}(18).\\
$\displaystyle p_{y}=-{\partial V\left(x,y\right)\over\partial y}\left[dt+{1\over v_{x}}\sqrt {\hbar\over 2m}dW\left(t\right)\right]$\hfill{}(19).\\
To describe the fluctuating wave directly, (7),(17) and (18) are needed.

The wave motion itself fluctuates because of the existence of this solution.\\
We expect that the result of the two-slit experiment can be explained using this new solution of Schr\"odinger equation: fluctuating wave motion.

For instance, there is an two-slit experiment with electron carried out by Tonomura et al$^{2}$. According to Tonomura et al this experiment is explained by the following theories.

\begin{center}
\textbf{4. The theory of two-slit experiment in advanced research 
laboratory, Hitachi Limited; Tonomura et al}$^{2}$\textbf{.}
\end{center}

The biprism consists of two parallel grounded plates with a fine filament 
between them, the latter having a positive potential relative to the former. 
The electrostatic potential is given by $V\left(x,z\right)$ and the incoming electron wave by $\exp i\left(k_{z}z\right)$, 
the deflected wave is given by

$\displaystyle \psi\left(x,z\right)=\exp i\left(k_{z}z-{me\over\hbar ^{2}k_{z}}\int_{-\infty}^{z}V\left(x,z'\right)dz'\right)$,\hfill{}T-(1)\\
The two waves having passed on each side of the filament can be approximated 
by $\exp i\left(k_{z}z\pm k_{x}x\right)$ up to a constant factor, where

$\displaystyle k_{x}=-{me\over\hbar ^{2}k_{z}}\int_{-\infty}^{\infty}\left(\partial V\left(x,z'\right)\over\partial x\right)_{x=a}dz'$,\hfill{}T-(2)\\
and the symmetry $V\left(x,z\right)=V\left(-x,z\right)$ has been taken into account.
This can be interpreted classically also:\\
$-e\left[\partial V\left(x,z'\right)/\partial x\right]_{x=a}$ is the $x$ component of force exerted on the electron. Its integral with respect 
to $dz/v_{z}=dt$ ($v_{z}=\hbar k_{z}/m$) gives the impulse imparted to it, which is the same in absolute value 
but reversed in sign, depending on which side of the filament the electron passes.
If the two waves overlap in the observation plane to give

$\psi\left(x,z\right)=\exp\left(ik_{z}z\right)\left[\exp\left(-ik_{x}x\right)+\exp\left(ik_{x}x\right)\right]$,\hfill{}T-(3)\\
then this leads to the interference fringes

$\left|\psi\left(x,z\right)\right|^{2}=4\cos^{2}k_{x}x$.\hfill{}T-(4)\\
The Copenhagen interpretation has concluded that $\psi\left(x,z\right)$ is the ``probability amplitude''.

\begin{center}
\textbf{5. Alternative explanation to Copenhagen interpretation: Waviness}
\end{center}

If $v_{z}$ is deleted from\\
$dz/v_{z}=dt$, $v_{z}=\hbar k_{z}/m$\\ 
that exists between the expression T-(2) and the expression T-(3) of the 
Tonomura thesis, it becomes $\displaystyle {m\over\hbar k_{z}}dz=dt$.\\ 
If this expression is substituted for the expression T-(1) of the Tonomura 
thesis, it becomes\\
$\displaystyle \psi\left(x,z\right)=\exp i\left(k_{z}z-{1\over\hbar }\int_{t_{0}}^{t}eV\left(x,z'\right)dt'\right)$.\\
And the expression T-(2) of the Tonomura thesis becomes\\
$\displaystyle k_{x}={p_{x}\over\hbar}=-{1\over\hbar}\int_{t_{0}}^{t} e\left(\partial V\left(x,z'\right)\over\partial x\right)_{x=a}dt'$.\\
However, because it is necessary to use the fluctuating wave motion (18),(19) accurately,
the impulse that electron receives is not\\
$-e\left[\partial V\left(x,z'\right)/\partial x\right]_{x=a}dt$\\
but\\
$\displaystyle -e\left[\partial V\left(x,z'\right)/\partial x\right]_{x=a}\left(dt+{1\over v_{z}}\sqrt{\hbar\over 2m}dW\left(t\right)\right)$.\\
(Note : While the direction of incidence is $x$ and the direction where electron that receives power bends is $y$ in (18),(19), the direction of incidence is $z$ and the direction where electron that receives power bends is $x$ in Tonomura thesis.)\\
This is the cause of fluctuation.

The momentum of electron fluctuates by this fluctuation of impulse, and the 
wave number vector of electron also fluctuates.

As understood when seeing (18),(19), the wave fluctuates only when it is in potential, and the wave doesn't fluctuate in the area that can be considered potential to be 0.

So, $k_{x}$ is a different value in each wave that has occurred 
from biprism.
Expression T-(2) means that $k_{x}$ is determined by the accumulation of the impulse 
that the electron wave received from potential energy when it went in the 
biprism. In general, because the impulse fluctuates, the impulse that 
the first wave received in the biprism and the impulse that the second wave 
received are different. As a result, the value of $k_{x}$ is different according to 
each wave.\\
As shown by the expression T-(3)\\
$\psi\left(x,z\right)=\exp\left(ik_{z}z\right)\left[\exp\left(-ik_{x}x\right)+\exp\left(ik_{x}x\right)\right]$\\
of the Tonomura thesis, there is $k_{x}$ in the wave function that is reaching the 
screen from right and left biprism. So the phase of wave fluctuates because 
it receives the influence of fluctuation of the impulse.

The appearance of the interference when wave number $k_{x}$ fluctuates is seen as 
follows.

By the expression T-(3) of the Tonomura thesis, the wave number vector of the 
wave that comes from the left can be written as $\left(k_{x},0,k_{z}\right)$ and the wave number vector 
that comes from the right, $\left(-k_{x},0,k_{z}\right)$ .

In general, because the impulse doesn't fluctuate symmetrically, 
neither $k_{x}$ from the left nor $k_{x}$ from the right are equal. Then, the wave number 
vector of the wave that comes from the left is written as $\left(k_{x}\left(L\right),0,k_{z}\right)$, and the wave 
number vector of the wave that comes from the right is written as $\left(-k_{x}\left(R\right),0,k_{z}\right)$. It 
becomes a different value because of the first wave, the second wave, and 
the third wave even if paying attention only to $k_{x}\left(L\right)$ ( paying attention only to $k_{x}\left(R\right)$ 
).

\begin{center}
Figure 1. When the wave number $k_{x}\left(L\right)$ that comes from the left and the wave number $k_{x}\left(R\right)$ that comes from the right are equal:
\end{center}
\includegraphics[width=120.000mm]{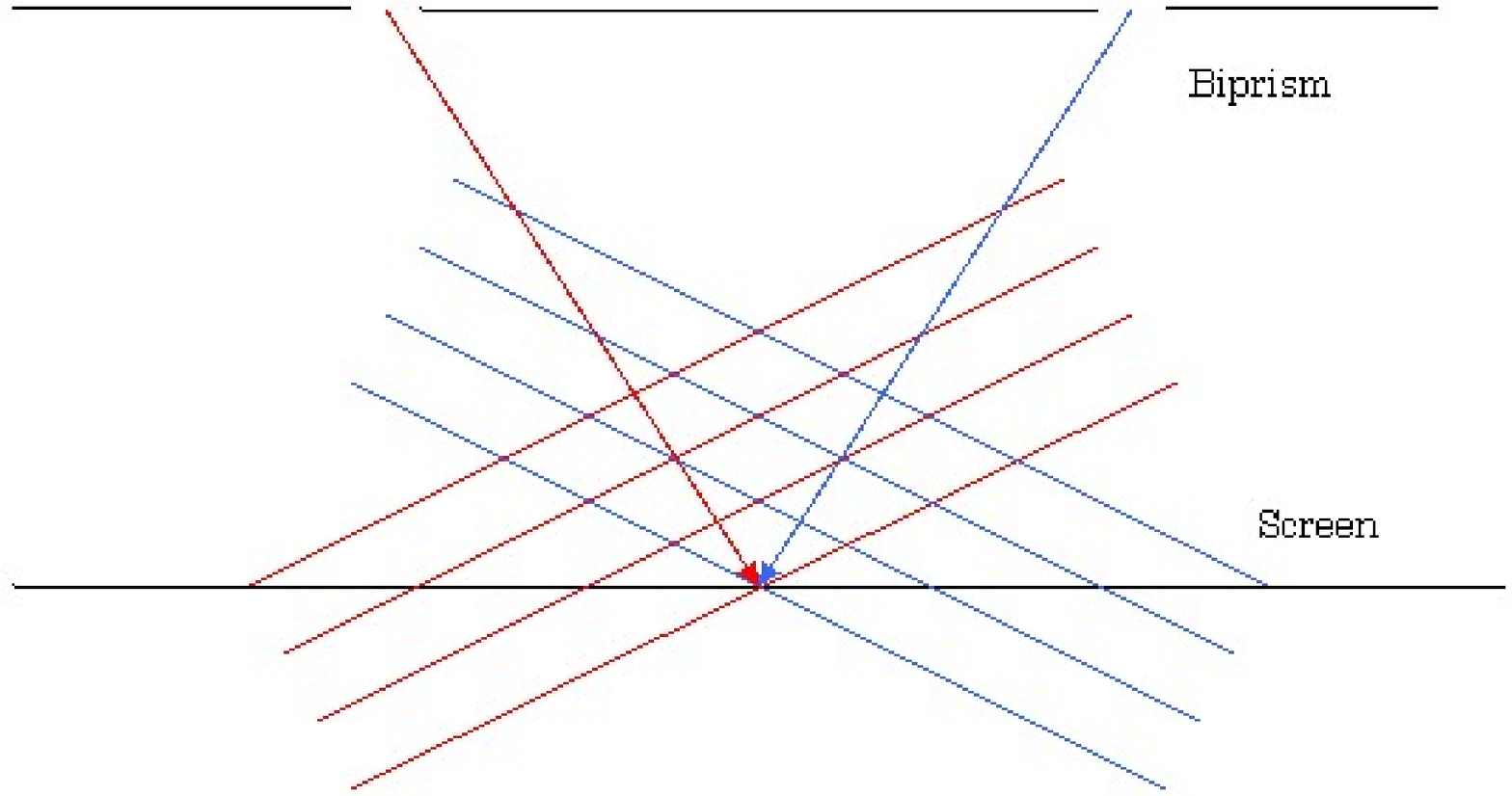}

Bright spot appears at the center of the screen.

\begin{center}
Figure 2: When the wave number $k_{x}\left(R\right)$ that comes from the right is larger than the Wave number $k_{x}\left(L\right)$ that comes from the left:
\end{center}
\includegraphics[width=120.000mm]{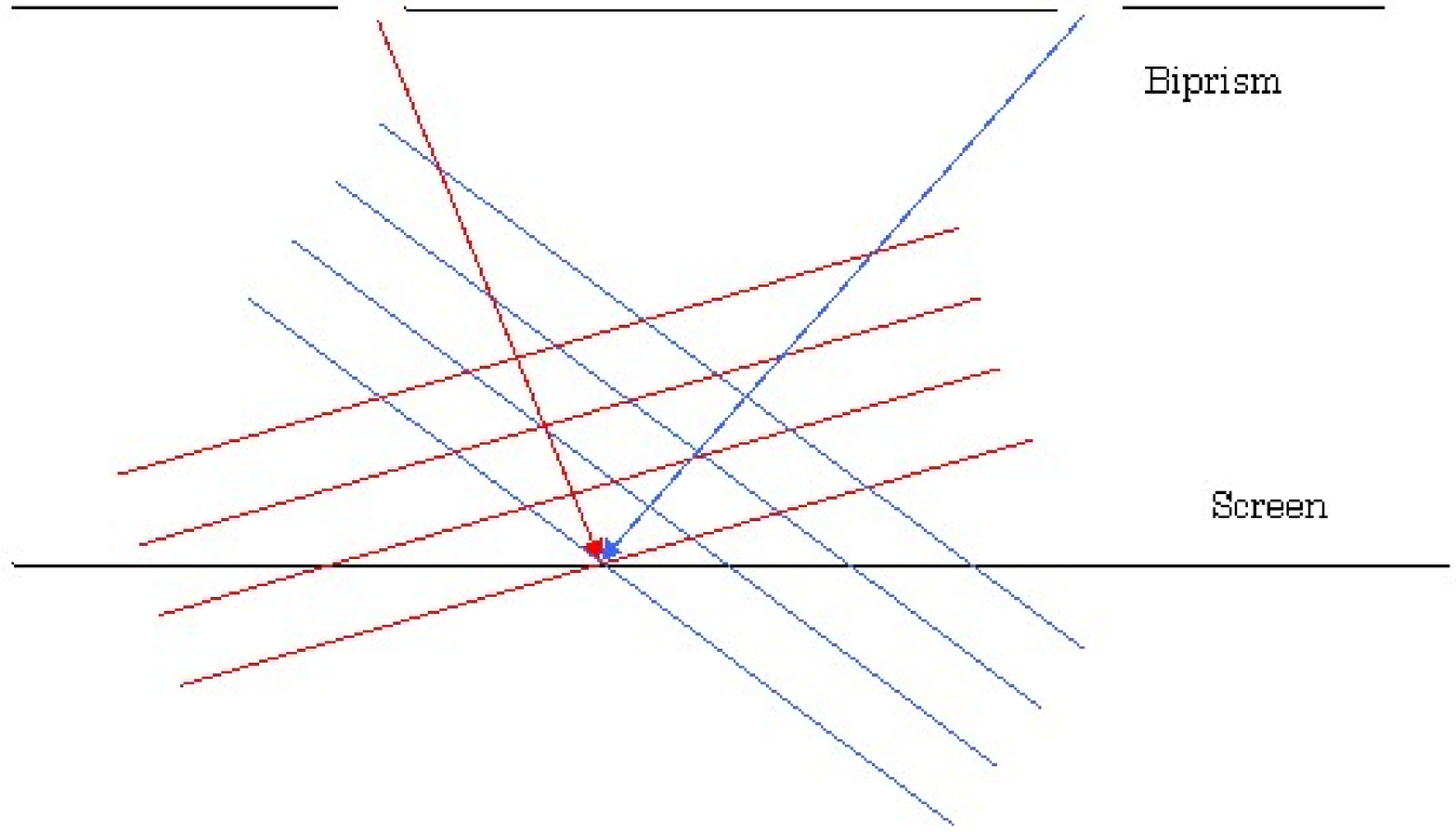}

Bright spot shifts left.

\begin{center}
Figure 3: When the wave number $k_{x}\left(R\right)$ that comes from the right is smaller than the Wave number $k_{x}\left(L\right)$ that comes from the left:
\end{center}
\includegraphics[width=120.000mm]{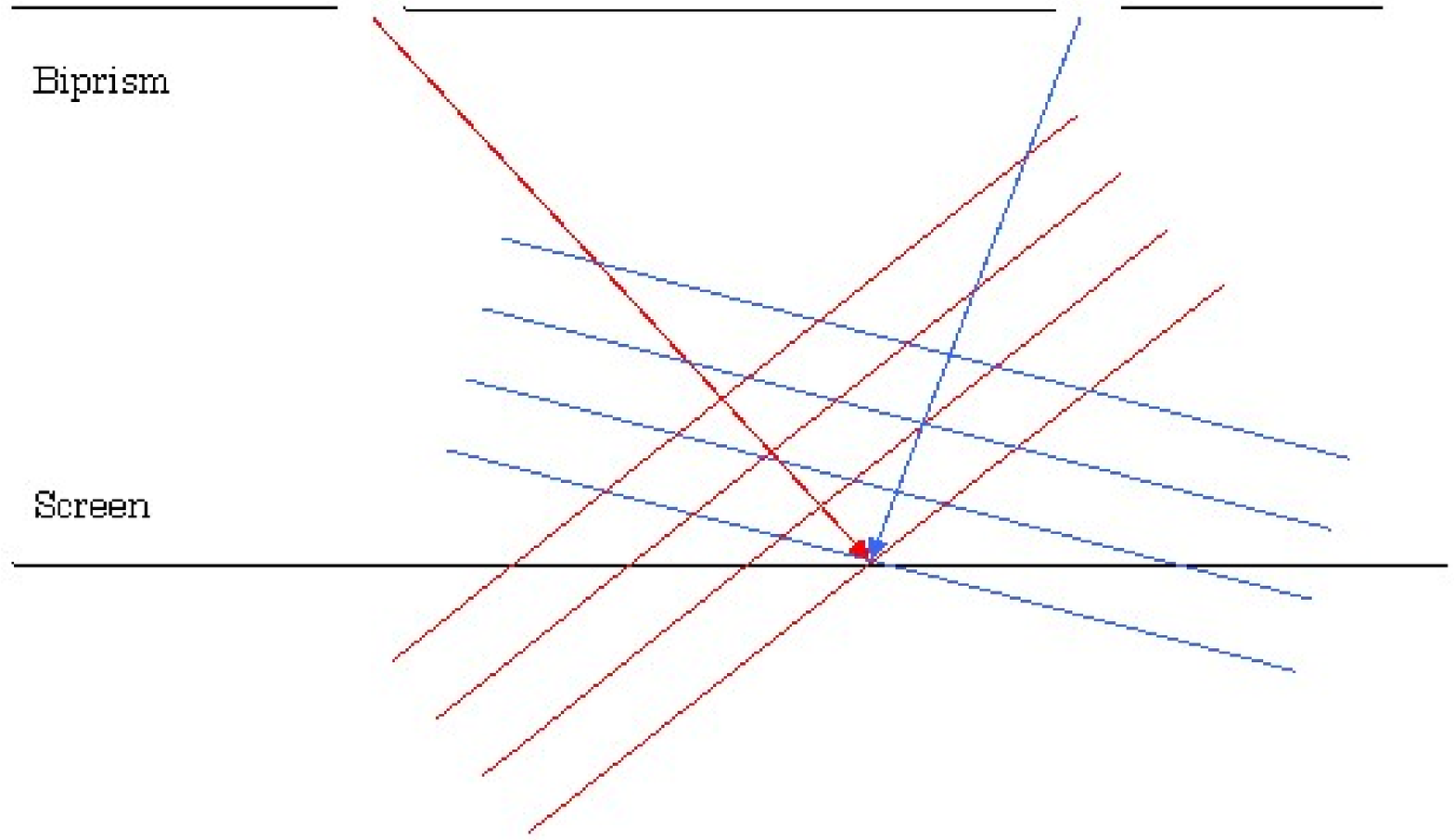}

Bright spot shifts right.
\hfill{}\\

We suggest that this is the mechanism by which the bright spot is observed 
at a random position. Because the impulse that the electron receives in the potential energy that the filament makes 
fluctuates, the wave number $k_{x}\left(L\right)$ of the wave that passes the left prism sometimes 
becomes large and at other times it becomes small. The wave number $k_{x}\left(R\right)$ of the 
wave that passes the right prism is also similar. As a result, the position 
that two waves strengthen each other is different on each occasion as shown 
in the above figure. In current quantum theory, only the case of Figure 1 is 
considered, and it is said that the place enclosed in the following figure 
is a position of the interference fringes.

\begin{center}
Figure 4: The position of the interference fringes:
\end{center}
\includegraphics[width=120.000mm]{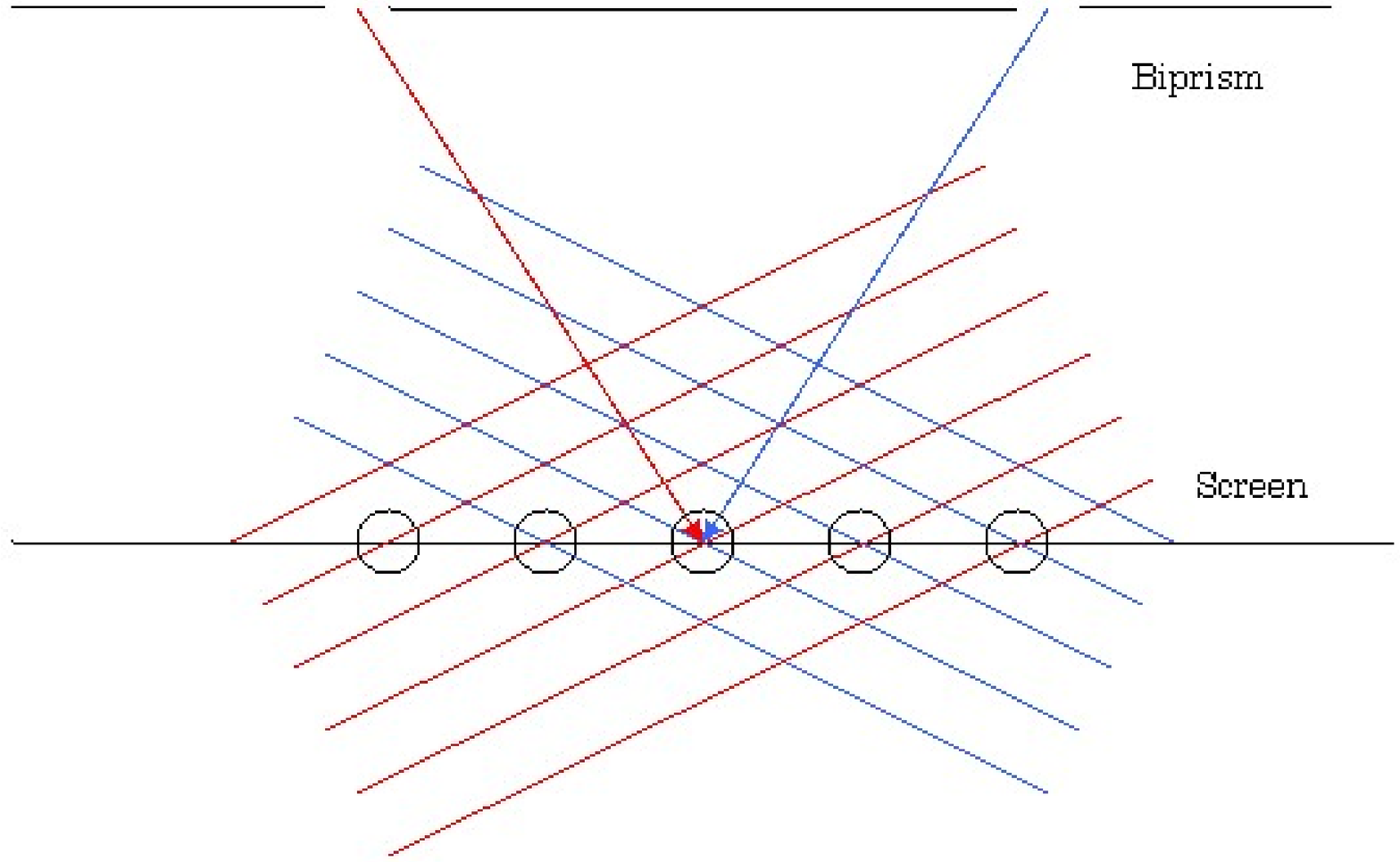}

It follows from this that our interpretation is different from the current 
interpretation of the quantum theory in that it becomes a very dynamic image 
like two moving searchlights independently scattering waves of light into 
the night sky. On the other hand, the image of current quantum theory is 
very static.

The second point that requires clarification is why it is observed as ``a 
spot'' in the experiment when the electron wave is weakened. The reason why 
it is observed as ``a spot'' is that the effect of the diffraction 
(so-called Fraunhofer diffraction) exacerbates the above-mentioned 
interference because the opening of biprism is not the ideal one like the 
delta function but has some size in an actual experiment. Therefore, 
strength of the electron wave on the screen becomes narrowed shape like the 
interference fringes shown by $\cos$ function narrowed by $sinc$ function ($\sin x/x$).

(Refer to the figure below. A part of the numerical value is excerpted from 
the Tonomura thesis.)

\begin{center}
Figure 5: The Fraunhofer diffraction in the two-slit experiment carried out by Tonomura et al:
\end{center}
\includegraphics[width=120.000mm]{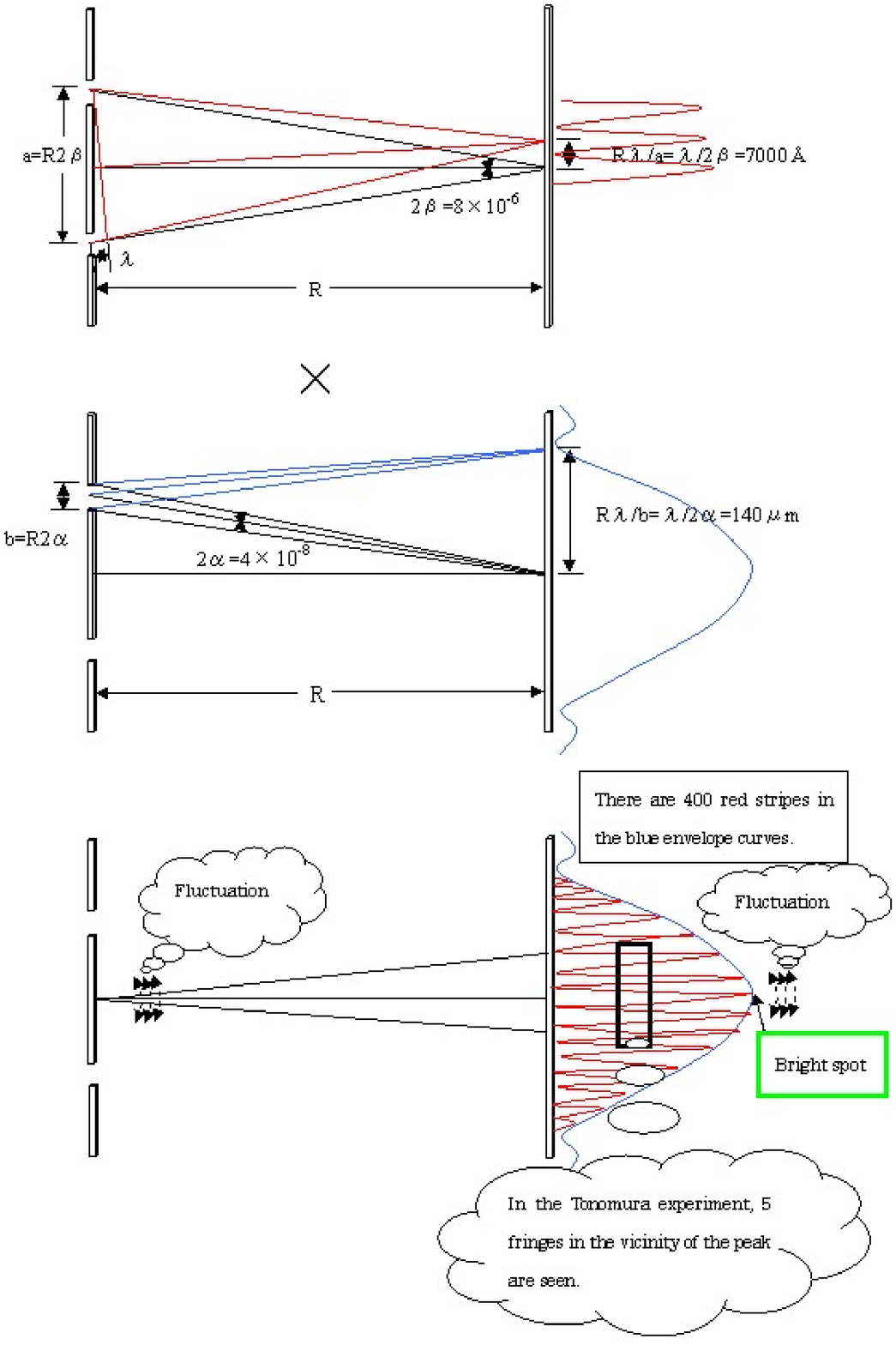}

Strength of the electron wave on the screen becomes shaped like a sliced 
mountain. There were an estimated 400 slices in the Tonomura experiment. In 
addition, only a very narrow area (center part of Airy disk so-called) in 
the top of a mountain will reflect because the pictures in this experiment 
were taken with very limited sensitivity. It is concluded that this is the 
bright spot observed on the screen.

Furthermore, the top of the mountain shakes at random due to the 
above-mentioned fluctuation. The peak of the distribution of the Fraunhofer 
diffraction appears at random because the potential energy (electric field) 
fluctuates and the electron wave fluctuates. Also, because the electron wave 
discharged from the electron gun is weak, only the part of the peak is taken 
of a picture.

Up to this point we have explained the two-slit experiment by only waviness.

\begin{center}
\textbf{6. Proposal to experiment}
\end{center}

According to Tonomura theses, the width of the interference fringes is $7000 \AA$ and the transverse coherence length is $140\mu m$. If the Fraunhofar diffraction pattern is a  probability wave said by a present quantum theory, the bright spot is sure to scatter over $280\mu m$, and to appear by many hundreds of (400 theoretically) interference fringes. It is because the electron reaches in the edge of the Fraunhofar diffraction  even though the probability is low.

On the other hand, the number of interference fringes is sure to be only ten or more in our proposal. It is because only peak of that figure is taken of picture, and it fluctuates.
The rough estimate is as follows.
The standard deviation (volatility) of fluctuating of the position of the peak of the Fraunhofar diffraction is\\
$\displaystyle \sigma=\sqrt{\hbar\over 2m}\sqrt{\Delta t}$\\
by (19).\\
According to Tonomura theses,\\ 
Distance from slit to screen : $1.5m$\\ 
Velocity of electron : $1.3 \times 10^8 m/s$ (Accelerating voltage : $50kV$).\\
So $\sigma$ becomes about $0.8\mu m$.\\
Therefore, ranges where the peak of the Fraunhofar diffraction is distributed are $2\times 0.8=1.6\mu m$ in $1\sigma$ (cover rate of $68\%$). ($\times 2$ is a meaning of both sides of normal distribution.)
Therefore, the number of interference fringes $=1.6\mu m \div 900\AA = 18$.
(Note : According to Tonomura theses, the width of the interference fringes is $900 \AA$ theoretically while it is $7000\AA$ experimentally. It is because a spherical wave instead of a plane wave is incident on the biprism in the actual experiment.)
So it is sure to be taken a picture of only ten or more interference fringes.
Moreover, because fluctuating of the peak of the Fraunhofar diffraction is normal distribution, it is expected that the interference fringe in the center part is bright, and it darkens to the surrounding.

\begin{center}
\textbf{7. Conclusion}
\end{center}

The wave motion itself fluctuates because of the existence of this solution\\
$\displaystyle \psi\left(x,t\right)=\exp\left(-{i\over \hbar}\int_{t_{0}}^{t}V\left(x,t'\right)dt'\right)\psi\left(x_{0},t_{0}\right)$\\
$\displaystyle +\exp\left(-{\pi\over 4}i\right)\int_{t_{0}}^{t}\exp\left(-{i\over \hbar}\int_{u}^{t}V\left(x,t'\right)dt'\right){\partial \psi\left(x,u\right) \over \partial x}\sqrt {\hbar\over m}dW\left(u\right)$.\\
We expect that the result of the two-slit experiment carried out by Tonomura et al can be explained as follows using this new solution of Schr\"odinger equation: fluctuating wave motion.

In the two-slit experiment, the wave number vector of each wave that occurs 
from 
biprism fluctuates by normal distribution. The wave that occurred from 
biprism is launched in various directions for this fluctuation. This 
fluctuation is expressed by the probability distribution of normal 
distribution.

In conclusion, we should note that while in present quantum mechanics it is 
the wave function that determines the probability distribution of the 
electron, our observations show that it is not the wave function but the 
kinetic energy exponential part that determines the probability 
distribution. As a result, wave motion itself fluctuates. Furthermore, the 
bright spot observed on the screen is not ``an electron'' but the peak of 
the distribution of the Fraunhofer diffraction.

\begin{center}
\textbf{Acknowledgements}
\end{center}

The author would like to thank Akitoshi Ishizaka for useful discussions about experiment. The author would like to thank Soei Koizumi and Denise Ochiai for their various supports and encouragement. The author also would like to thank Yasuhiro Funamizu and Tamotsu Saito for their warm encouragement and support. The author wishes to thank Jun Kawaguchi, Yoshio Moriya and Yoshitaka Sato. The author wishes to thank Naomi Matsumoto for kind support.

\begin{center}
\textbf{Appendix A ``Normal distribution and kinetic energy''}
\end{center}

In general, when $x$ follows the Stochastic process\\
$\displaystyle x-x_{0}=dx=\mu dt+\sigma dW=\mu dt + \sigma \sqrt{dt}\xi$\\
($dW$ : Standard Brownian motion, $\xi$ : Standard regular random variable), $x$ becomes normal distribution\\
$\displaystyle \Phi\left(x,t\right)={1\over\sqrt{2\pi\sigma^{2}dt}}\exp\left[-{(x-x_{0}-\mu dt)^{2}\over 2\sigma^{2}dt}\right]$\hfill{}(20).\\
On the other hand, a kinetic energy paragraph of path integral becomes\\
$\displaystyle \sqrt{m\over 2\pi\hbar dt}\exp\left[-{1\over \hbar}{1 \over 2}m\left({dx \over dt}\right)^2 dt\right]$\\
after Euclidean approach.\\
$\displaystyle \sqrt{m\over 2\pi\hbar dt}\exp\left[-{1\over \hbar}{1 \over 2}m\left({dx \over dt}\right)^2 dt\right]=\sqrt{m\over 2\pi\hbar dt}\exp\left[-{1\over \hbar}{1 \over 2}m{\left(dx\right)^2 \over dt} \right]=\sqrt{m\over 2\pi\hbar dt}\exp\left[-{1\over \hbar}{1 \over 2}m{\left(x-x_{0}\right)^2 \over dt} \right]$\hfill{}(21),\\
So, (20) corresponds to (21) after replacing\\
$\displaystyle \sigma=\sqrt{\hbar\over m}$, $\mu=0$.\\
In a word, the process $x$ follows is not\\
$\displaystyle dx=\mu dt+\sigma dW$\\
but\\
$\displaystyle dx=\sigma dW$\\
in path integral.

\begin{center}
\textbf{Appendix B ``Ito's lemma''}
\end{center}

When $X$  follows the Ito process\\
$\displaystyle dX=a\left(X,t\right)dt+b\left(X,t\right)dW\left(t\right)$,\\
the movement of function $f\left(X,t\right)$ of $X$ and $t$ follows\\
$\displaystyle df=\left(a\left(X,t\right){\partial f\over \partial X}+{\partial f\over \partial t}+{1 \over 2}{\partial^2 f\over \partial X^2}\left\{b\left(X,t\right)
\right\}^2 \right)dt+{\partial f\over \partial X}b\left(X,t\right)dW\left(t\right)$.\\
Here, $dW\left(t\right)$  means Standard Brownian motion.

\begin{center}
\textbf{References}
\end{center}

\noindent $^{1}$M. Kijima, Stochastic processes with applications to finance, [Chapman {\&} Hall/CRC, 209-210 (2002)]\\
$^{2}$A. Tonomura, J. Endo, T. Matsuda, T. Kawasaki, and H. Ezawa, Demonstration of single-electron buildup of an interference 
pattern, [American 
Journal of Physics 57, 117-120 (1989)]\\

\end{document}